\documentclass{llncs}

\usepackage{epsfig}

\begin{document}

\title{Using LDAP Directories for Management of PKI Processes}

\author{
Vangelis Karatsiolis\inst{1,}\inst{2}, Marcus Lippert\inst{1} \and Alexander Wiesmaier\inst{1}
}

\institute{Technische Universit\"at Darmstadt,
           Department of Computer Science, \\
           Hochschulrstra\ss e 10, D-64289 Darmstadt, Germany \\
           \email{\tt {karatsio,mal,wiesmaie}@cdc.informatik.tu-darmstadt.de}
\and
           Fraunhofer Institute for Secure Telecooperation, \\
           Dolivostr. 15, D-64293 Darmstadt, Germany \\
           \email{\tt karatsio@sit.fhg.de}
}

\maketitle

\begin{abstract}

We present a framework for extending the functionality of LDAP servers from their typical
use as a public directory in public key infrastructures. In this framework the LDAP servers are
used for administrating infrastructure processes. One application of this framework
is a method for providing proof-of-possession,
especially in the case of encryption keys. Another one is
the secure delivery of software personal security environments.

\bigskip

\end{abstract}

\section{Introduction}

A Public Key Infrastructure (PKI) provides practical use of public
key cryptography proposed by Diffie and Hellmann \cite{DH76}.
Entities register for the services of the PKI in order for them to
achieve security goals like authenticity and confidentiality using
digital signatures and encryption. One of the most typical
products of a PKI are digital certificates. They bind a public key
to an entity which owns\footnote{An entity is considered to be the
owner of a private key if and only if it can legally use it.} the
corresponding private key. Commonly met certificates follow the
X.509 \cite{X.509} and the RFC 3280 \cite{RFC3280} standards.
A PKI is usually composed of two components namely the Registration
Authority (RA) and the Certification Authority (CA).

The task of the RA is to register an end-entity in the PKI by
examining its credentials. This entity may either already possess
a key pair or not. In the first case, the public key is provided
by the entity to the RA during registration and forwarded to the
CA that issues the certificate. In the second case, the key pair
is produced by the CA and, together with the issued certificate,
delivered to the entity. In either case, the CA has to guarantee
that finally the certificate is issued for the entity that owns
the corresponding private key.

The problem that arises in the first case is for the CA to
determine whether this entity does really own the corresponding
private key or not. Therefore, the entity must provide a
Proof-of-Possession (PoP) for this private key. In several
certification request syntaxes special fields are reserved for
this purpose. An important issue with PoP is the number of
messages needed for accomplishing it. While for signature keys the
straightforward solution is to provide a signature during
registration, the situation is not so easy for encryption keys.
One good solution in terms of number of messages is the indirect
scheme where the issued certificate is encrypted with the included
public key and thus can only be read by the intended recipient. We
provide an easy and efficient variant of this scheme using a
directory based on the lightweight directory access protocol
(LDAP) \cite{RFC3377}. Furthermore, it can be extended to tightly
link the PoP to the registration procedure. Moreover, a successful
PoP is necessary for activation of the certificate. Whether or when
this activation will take place is under full control of the entity.

In the second case, the problem is to securely deliver the private
key to the registered and identified entity. A commonly used
standard for software personal security environments (PSE) is
PKCS\#12 \cite{PKCS12}. Usually, such PSEs are either handed out
face to face or sent by e-mail. For automated certification
processes in which management should be simple, the solution of
physical presence can not be used since it hampers the automated
process and requires human administration. Sending the PKCS\#12
file by e-mail is secure since PKCS\#12 has mechanisms to ensure
privacy and integrity. Nevertheless, the e-mail may be intercepted
by a third party which can try to extract the keys (in a password
protected PKCS\#12 file this may be easy for a weak password).
Moreover, the e-mail may never reach his recipient due to an incorrectly
configured spam filter or even a mailbox which has reach
its capacity limit. We suggest a scheme based on an LDAP directory
to ensure that only the legitimate recipient will receive his PSE
without any eavesdropping from a third party.

\bigskip

This paper is organized as follows: In Section
\ref{sec:messages}, we discuss specifications on certification request messages
and examine their proposals for achieving PoP. In Section \ref{sec:schemes}, we describe
our proposed schemes and argue on their usability. Lastly, in Section \ref{sec:conclusion}
we conclude the paper and discuss future work in this direction.

\section{Certification Request Messages and PoP}
\label{sec:messages}

Common practice for the registration of end-entities in a PKI
environment are special certification request message syntaxes.
Several specifications address this problem.

The most commonly used one is the PKCS\#10 \cite{PKCS10}. It
defines a syntax in which entities can provide information needed
for creating a certificate from the CA. Furthermore, it enables
them to include other data that can be used during and after the
certification process. This syntax requires the message to be
secured with a digital signature. Among other purposes, this is
done to provide a PoP for the corresponding private key, since the
signature can be verified from the receiving party (RA or CA).
However, this syntax does not consider keys used for encryption
only. Such keys might have to be treated differently due to
limitations of the cryptosystem or different security
requirements.

Another specification is the Certificate Request Message Format
(CRMF) \cite{RFC2511}. While handling signature keys identically
to PKCS\#10, it provides three methods for PoP of encryption keys.
The first is to reveal the private key to the CA. The second one,
called direct method, requires exchange of challenge-response
messages and thus needs extra messages. The third one, the
indirect method, is to encrypt the issued certificate with the
contained public key and have the end-entity to send the decrypted
certificate back in a confirmation message and by this demonstrate
ownership of the corresponding private key.

Similar proposals can be found in Certificate Management Protocol
(CMP) \cite{RFC2510}. Special care about encryption keys is taken
also in Certificate Management Messages over CMS (CMC)
\cite{RFC2797}. CMC specifies a mechanism for PoP which requires
more than a single-round trip\footnote{A request which is done from the
end-entity, processed from the CA and returned to the end-entity.}
and therefore complicates the process. Lastly, XKMS \cite{XKMS}
considers the PoP problem exclusively for signature keys. For a
discussion on PoP see also \cite{ANL03}.

The above proposals do not provide satisfactory solutions in cases
where the end-entity does not want to reveal its key or the
certification process management should be kept minimal and
simple.

We believe that PoP is of great importance in the case of
encryption keys. A certificate containing a public key for which
the corresponding private key is not owned by the intended
end-entity or does not exist at all is actually unusable. Any
message encrypted with this public key can not be decrypted by the
end-entity. This scenario can become extremely dangerous in cases
where the original message is destroyed\footnote{This is a typical
function of many encryption clients.} and only the encrypted
message exists.
Therefore, no one should be able to use a
certificate containing encryption keys without prior
acknowledgement of its owner.

To solve the PoP problem we propose a scheme similar to the
indirect method described above, in which an end-entity is an
authenticated user of an LDAP directory, which he can download and
activate his certificate from. But in order to do so, he must
decrypt a secret (implicit use of his private key), decrypt the
certificate and then put it back on the LDAP server. Thus, we can avoid
extra confirmation messages to the CA.

\section{The New Schemes}
\label{sec:schemes}

\subsection{LDAP basics}

LDAP is a protocol for providing access to directories based on
the X.500 specification \cite{X.500}. The current version is
LDAPv3 specified in \cite{RFC3377}. Internet directories speaking
LDAP are most commonly used in PKIs for dissemination of public
information. It is exactly this ``public directory'' mentioned in
\cite{DH76} when public key cryptography was discovered. LDAP
directories serve as the place where clients can download
certificates of other users in order to send encrypted messages or
verify digital signatures. In addition they can be informed with
the latest certificate revocation information by downloading a
certificate revocation list (CRL) from the directory. Therefore,
almost all PKIs support LDAP and no significant rearrangements
should be done from their side to use an LDAP directory for
managing processes in their environments.

Apart from the fact that LDAP is already supported by PKIs, it
has also attractive security features. It provides simple authentication
of users with a password\footnote{When used alone it does not
really provide significant security since the password travels in
clear.}. In addition, it supports the simple authentication and
security layer (SASL) \cite{RFC2222} as illustrated in
\cite{RFC2829}. Lastly, it can be combined with TLS \cite{RFC2246}
as described in \cite{RFC2830}. The communication over TLS
guarantees privacy and integrity of the data exchanged between an
LDAP directory and a client. The password authentication can be
used in combination with TLS to avoid that the password is
transmitted in clear. For a thorough study on the security of
directories see \cite{Cha00}.

\subsection{Providing PoP with LDAP}

We present the scheme for providing PoP for encryption keys using
an LDAP directory.

In this scheme the CA is accepting a certification request for
encryption keys. At this point the CA assumes that the private key
exists and issues the certificate. After that, it creates an entry
on the LDAP directory as usual. One difference is that the
certificate is encrypted with the public key it contains and
stored in the attribute \textit{encryptedUserCertificate} (see Appendix A). 
Secondly, the CA randomly chooses an LDAP user
password. Its hashed value is placed in the entry's attribute
\textit{userPassword} for the simple authentication procedure of the LDAP. It also
encrypts the password using the entity's public key and puts it in
the special attribute \textit{encryptedUserPassword}. At this point even
the user itself does not know his password, since only its hash
value and an encrypted version is found on the directory.

\begin{figure}[t]
\centerline{\epsfig{figure=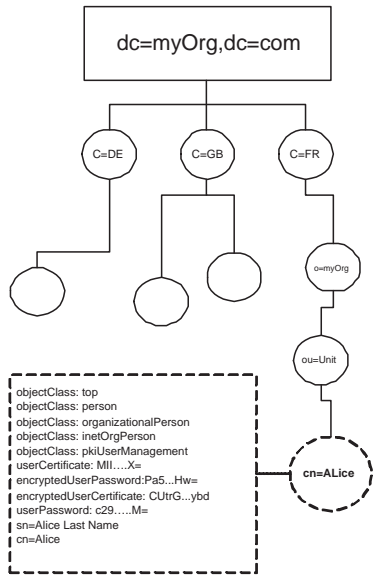,height = 3in}}
\caption{\label{fig:DIT} End-user entry in an LDAP server.}
\end{figure}

We have introduced a new object class called \textit{pkiUserManagement} (see
Appendix A) which can hold the above described attributes. Figure
\ref{fig:DIT} shows an example of such an LDAP entry along with
its attributes.

The end-entity, that owns the corresponding private key, does
the following: First, it downloads the encrypted password and
certificate and decrypts them with its private key. Now it can use
the password to bind to the directory as an authenticated user and
write the decrypted certificate in its entry. With this the PoP is
completed and, simultaneously, the certificate is activated
enabling other entities to use it. On the contrary, for an
end-entity that does not own the private key, both the certificate
and the password will remain encrypted in the directory and cannot
be used. Clearly, the communication between the authenticated user
and the LDAP server must be secured with SASL or TLS to avoid
transmitting the password in clear.

One step that the CA must perform is to force the LDAP directory,
by proper access control lists (ACLs), to accept write requests
only for authenticated users and only for the \textit{userCertificate}
attribute of their own entry. This enforces that exclusively authenticated users can
place their certificate only on their own entry
in the directory. In addition, these ACLs must be configured in
such a way that the user password (although hashed) is not visible
to other users. This will eliminate the possibility of dictionary
attacks on the password performed from any end-user.

An optional variant of this scheme, which can link both identity
and PoP information more tightly together, is the following:
The encrypted password (when decrypted) is only the
half of the real password\footnote{Which is found hashed to the directory.}
used by the user to authenticate to the LDAP
directory. The other half of the password is provided
to the user during registration. This half works as a shared-secret between the CA and the
end-user. The user, in order to authenticate to the directory, has to combine the two passwords.

A third variant is to have the shared-secret
function as the (complete) password for the user to authenticate to the directory. 
PoP will follow with decryption of the certificate without having to decrypt the
password too. Nevertheless, in this case the ability to authenticate to
the directory is disconnected from PoP.

CAs can realise this scheme differently based on their policies.
For example, if after three days the certificate is still
encrypted in the directory, the user is deleted from the
directory leaving him no ability to use this certificate. Alternatively, after
decryption of the certificate the user password is deleted in
order for the user to be unable to authenticate to the directory.

\subsection{Methods overview}

We discuss the characteristics and properties of the certification request messages
presented in Section \ref{sec:messages} and our proposed LDAP based solution. PKCS\#10
and XKMS do not support PoP for encryption keys (Encryption Key Support, EKS) and we will
not discuss them further in this section. CMC and the LDAP
based proposal specify only one possible method to provide PoP, while CRMF
and CMP offer three different ones (Different Methods, DM). Nevertheless, the last ones
propose revealing the
private key (Private Key Revealing, PKR) as a solution. Furthermore, CMC requires
more than a single-round trip where all other methods do not. But CMP
and CRMF require that a confirmation message should be 
sent to the CA. Our proposed method does not require those (Single Round Trip, SRT).
In table \ref{tab:ComparisonTable} we can see an overview of these characteristics.

\begin{table}[t]
\begin{center}
\begin{tabular}{|l||c|c|c|c|c|c|}
\hline
         & CRMF & CMP & CMC & Proposed Solution \\
\hline
EKS      & +   & +  & +  & + \\
\hline
DM       & +   & +  &  -   &  -  \\
\hline
PKR      & -   & -  &  +   & +   \\
\hline
SRT      &  -   &  -   &  -   & + \\
\hline
\end{tabular}
\end{center}
\caption{\label{tab:ComparisonTable} Comparison Table.}
\end{table}

\subsection{Secure delivery of software PSEs with LDAP}

A variation of the above scheme can be used for delivering
software PSEs (usually in the PKCS\#12 format).
In this scenario the PKCS\#12 structure is placed in the \textit{userPKCS12} \cite{RFC2798}
attribute of the LDAP entry. The connections done to the directory must be secured
with TLS. If a user can authenticate to the directory, then he has the ability
to download its own PSE. But in order to authenticate, he needs a password
provided during registration. Therefore only the intended user can access his PSE.

To enforce this, the CA has to choose special ACLs. Only the authenticated users 
must have read access to the \textit{userPKCS12} attribute of their own entry, while all
others should have no privileges at all. Downloading the PSE is performed with integrity
and confidentiality due to the TLS connection.

We have implemented both schemes to provide a proof of concept. We have used
the OpenLDAP \cite{OpenLDAP} directory since it gives the possibility to
create flexible ACLs and supports TLS and SASL. We have used JNDI \cite{JNDI}
at the CA and client side for the LDAP interfaces.

\subsection{Usability issues}

We believe that these two basic applications for user management
with LDAP enhances the usability of the PKI. In the PoP case it is
easier for a human user to decrypt a value than to provide his
private key or to engage himself in challenge-response procedures.
Decrypting data like the certificate or the password is the
intended usage of the decryption key and is already implemented in
common client software. Furthermore, most client software comes
with LDAP support to be able to fetch the certificates of others.
These features can be easily extended and combined to
automatically fetch the encrypted data from the directory, decrypt
it and store the certificate back. Particularly, the last action
can be totally transparent to the end-entity.

Also, fetching a PSE from the directory is essentially the same as
looking up certificates of others. Thus, no extra development
needs to be done for the client software. The usability is
increased by the fact, that the client can transparently
download the PKCS\#12 file and integrate the contained private
key at the client side\footnote{Probably asking for an extra password for
its internal database.}. Instead, if the user has received his PSE
by e-mail or floppy disc, he would have to install it
somehow manually. Furthermore, only the LDAP connection has to be
configured at the client and nothing else. Depending on the CA's
policy, also the following strategy may be chosen: The software
PSE may additionally remain on the directory to serve as backup.
Moreover, it can be available from everywhere. For this, client software could be
configured to download and use the PSE on demand without mandatory storing it locally.

\section{Conclusion}
\label{sec:conclusion}

We have presented two schemes for providing proof-of-possession
and secure delivery of personal security environments.
Both schemes use an LDAP directory which is already integrated in most PKIs,
it is ideal for user management, it has sufficient security features and
many existent clients have already LDAP interfaces which can
be extended. We plan to integrate our proposed solutions into an upcoming PKI
installation at the University of Darmstadt. Future work will
concentrate on extending the use of LDAP
in PKIs for other administration and certificate management
tasks like certificate revocation and certificate renewal. 


\section*{Appendix A}\label{app:A}

Object Classes \\

( 1.3.6.1.4.1.8301.3.2.2.1.6 NAME 'pkiUserManagement' SUP top 

AUXILIARY MAY ( userEncryptedPassword \$ userEncryptedCertificate ) ) \\

\noindent
Attributes\\

( 1.3.6.1.4.1.8301.3.2.2.1.7 NAME 'userEncryptedPassword' 

EQUALITY octetStringMatch 

SYNTAX 1.3.6.1.4.1.1466.115.121.1.40 SINGLE-VALUE ) \\

( 1.3.6.1.4.1.8301.3.2.2.1.8 NAME 'userEncryptedCertificate' 

EQUALITY octetStringMatch 

SYNTAX 1.3.6.1.4.1.1466.115.121.1.40 SINGLE-VALUE )

\end{document}